# Ultracompact Nano-Mechanical Plasmonic Phase Modulators

B. S. Dennis[1], M. I. Haftel[2], D. A. Czaplewski[3], D. Lopez[3], G. Blumberg[1] and V. Aksyuk[4#]

[1]Rutgers, the State University of New Jersey, Dept. of Astronomy and Physics, Piscataway, NJ 08854, USA, [2]University of Colorado, Dept. of Physics, Colorado Springs, CO 80918, USA, [3]Argonne National Laboratory, Center for Nanoscale Materials, Argonne, IL 60439, [4]Center for Nanoscale Science and Technology, National Institute of Science and Technology, Gaithersburg, MD 20899, USA. [#]e-mail: vladimir.aksyuk@nist.gov

**Dielectrics' refractive index limits photonics miniaturization. By coupling light to metal's free electrons, plasmonic devices[1,2] achieve deeper localization[3-7], which scales with the device geometric size. However, when localization approaches the skin depth, energy shifts from the dielectric into the metal, hindering active modulation. Here we propose a *nano-electromechanical* phase modulation principle exploiting the extraordinarily strong dependence of the phase velocity of metal-insulator-metal(MIM)[8] gap plasmons on dynamically variable gap size.[9,10] We demonstrate a 23μm long non-resonant modulator having 1.5π rad range with 1.7dB excess loss at 780nm. Analysis shows an ultracompact 1μm$^2$-footprint π rad phase modulator can be realized, more than an order of magnitude smaller than any previously shown. Remarkably, this size reduction is achieved without incurring extra loss, since the nanobeam-plasmon coupling strength increases at a similar rate as the loss. Such small, high density electrically controllable components may find applications in optical switch fabrics[11] and reconfigurable flat plasmonic optics[12].**

Phase modulators, often used as active elements in photonic switches, enable flexible provisioning of communication channels and reconfiguration of networks at the physical layer. The application requirements for switches are distinct from those of data modulators: switching can be slower than data modulation rates, with a premium put on compactness, low power consumption, wide optical bandwidth and low optical losses. As nanophotonic optical communication architectures and technologies are being developed in response to inter-chip and on-chip electronic bottlenecks, more compact, low-power optical switch fabrics, with 1 μs to 10 ns switching times, would enable new functionality, such as flexible signal routing and dynamic reconfiguration of the optical layer, architecturally analogous to electronic field-programmable gate arrays.

Several different modulation principles have been proposed and used to realize a variety of compact phase modulators. Because most of them are aimed at data modulation, none directly explore the ultimate limits of size scaling, with the smallest footprint ≈ 30 μm$^2$. The non-resonant devices have limited phase modulation strength per area and include thermo-optical devices[13] with large power dissipation, very fast slot plasmon electro-optical devices,[14] where device size is limited by the Pockels effect, and electro-mechanical devices.[15] Resonant electro-optical[16] and electro-mechanical[17,18] devices achieve higher phase-modulation strength at the expense of reduced optical bandwidth. Carrier injection based semiconductor and plasmonic devices[7,19] tend to have large absorption modulation, that results in high excess loss for phase modulation. These work well as intensity modulators, which are not suitable for realizing passive 1x2, 1xN or NxN switch fabrics.

In this letter, we propose and demonstrate a gap plasmon phase modulator (GPPM), where we essentially create a plasmonic metamaterial with a large effective Kerr coefficient driven by electromechanical geometric reconfiguration. Therefore the refractive index for in-plane gap plasmon (GP) modes varies strongly with applied voltage. GPs are broadband optical propagating modes[8,10] that can be vertically and laterally confined to sub 100 nm gaps between two metal layers, forming some of the smallest known optical waveguides and resulting in significant field enhancements,[3,5,20,21] Low loss coupling into such small GP waveguides has been demonstrated[5], making possible efficient connections to conventional dielectric waveguides for long distance interconnects.

The GPPM exploits the high sensitivity of the GP phase velocity to changes in the gap size by making one of the metal layers mechanically moveable via electrostatic actuation. No optical resonator is used to enhance the phase modulation and there is no low-frequency guided mode cutoff,[22] making the modulation principle optically broadband, capable of operating from the visible to the far-infrared.[9,23-25] The phase modulation strength per area of our experimentally demonstrated GPPM is comparable to that of resonant devices and an order of magnitude better than that achieved in mechanically-tuned dielectric slot waveguides.[15] While in dielectric slots the effective index tends to a fixed value as the gap is reduced, in GP it continues to increase steeply, underlying the unique GPPM scalability.

The nature of confined energy modes at optical frequencies in plasmonic devices can itself be understood as electro-mechanical[26] as opposed to electro-magnetic, with the kinetic energy of electrons, together with the Coulomb energy, playing a critical role and enabling localization at much smaller scales. This confinement comes at the expense of increased losses through inelastic electron scattering, which may impose fundamental limitations on scaling of any plasmonic device and thus should be thoroughly understood. We present an analytical investigation showing, remarkably, that our GPPMs can be scaled down by at least a factor of 100 in area to a sub 1 $\mu m^2$ footprint, while maintaining the $> \pi$ rad modulation depth and $\approx 5$ dB optical loss. The optomechanical modulation strength increases with decreasing gap, and the propagation losses can be kept constant by shortening the device length.

Our GPPM is based on an electrostatically tunable gold-air-gold waveguide fabricated from a gold-$SiO_2$-gold MIM stack (see Fig. 1 and Methods) with initial air gap around 280 nm. The top gold film is patterned into suspended deformable metal beams $23.0 \pm 0.5$ $\mu m$ in length and $1.50 \pm 0.07$ $\mu m$ wide supported at both ends by $SiO_2$ pillars with the GP propagating underneath and along the beams. When a voltage is applied, the electrostatic force deforms the beams down into an approximately parabolic shape, narrowing the MIM gap at the beam center by about 80 nm as the applied voltage increases from 0 V to 7 V (Fig. 1 c-e).

To measure the GPPM optical performance, a 780 nm wavelength Gaussian laser beam is focused from above onto an in-coupler grating cut into the top film at the GPPM input, launching a Gaussian GP mode into the device. An out-coupler slit at the output of the GPPM is used to sample the modulated plasmon using a microscope from below. A window in the top gold film above the out-coupler slit allows introduction of a tilted-reference optical beam for phase-sensitive imaging of the modulated plasmon. Using a Mach-Zehnder type interferometer with the reference split off from the excitation laser, both the GP phase retardation and optical loss are measured as a function of waveguide gap by electrostatically controlling the GPPM beam displacements (see Fig. 1 and Supplementary). We collect optical micrographs of the out-coupled GP light with and without the reference optical beam at different applied DC voltages. The interference and GP-only intensity profiles are extracted by integrating the micrograph data in the direction normal to the slit (see Fig 2a, top for a representative interference micrograph).

Fig. 2 shows these profiles at different applied voltages for one of the devices. The interference patterns shift to the right as the GP phase is retarded with the increased voltage, while in the absence of the reference beam the main change is a slight decrease in the plasmon intensity. The GP-only profile shapes (Fig. 2b) are fit by a common Gaussian profile with intensity as the only adjustable parameter for each voltage, as they are created by the Gaussian excitation beam focused on the in-coupler grating. The interference profiles (Fig. 2a) are fit well by the expected interference pattern[28] between the known, common reference Gaussian beam and the measured GP-only intensity data for the particular voltage from Fig 2b. The good agreement indicates that the GP remains collimated with a flat wavefront and no phase distortion is introduced by the GPPM. The GP phase relative to the reference beam is the only adjustable parameter for each profile fit, while the intensity of the reference Gaussian is a single extra adjustable parameter that is common for all the interference micrographs for a given device. All other parameters including the reference Gaussian width, center and the wavefront angle are separately measured and fixed (see Methods).

Fig. 3a shows the phase change induced by the GPPM with 0.0 V to 7.0 V applied, as a function of the gap, which is measured at the narrowest point at the center of the device. The excess optical power loss, caused by the narrowed gap under the actuated beams can be seen in Fig. 3b, which plots the integrated areas of Gaussian intensity fits from Fig. 2b normalized by that of the unactuated device. A phase shift exceeding 5 rad is achieved, while the corresponding excess loss is near 30 % (1.7 dB) (gold data points in Fig. 3) when the gap is tuned by approximately 30 %, from 270 nm to 190 nm.

The GPPM has an optomechanical modulation strength of 54 mrad/nm ± 4 mrad/nm, producing $\frac{3\pi}{2}$ rad phase shift, which can be compared to the $\frac{\pi}{2}$ rad shift demonstrated in a 170 μm long optomechanical dielectric device.[15] A modulation range in excess of π radians is required by many practical switching and modulation applications. To understand the GPPM performance, we developed a semi-analytical model of one dimensional GP propagation [Supplementary]. The analytic results of GP phase shift and intensity calculations agree well with measured data (Fig. 3a, b solid line). The calculated intrinsic insertion loss, through an unactuated device is 5.3 dB.

Unlike dielectric waveguides, MIM waveguides support a guided mode for any frequency below the surface plasmon (SP) resonance and for gaps down to the single nanometer range, where below that, local classical theory begins to break down.[29] The effective index increases and the GP wavelength decreases dramatically in small gaps.[3,9] Moreover, the strength of the phase modulation in this geometry increases (inset of Fig. 3a) approximately inversely with the square of the gap, g, so that $d\phi/dg \sim 1/g^2$, in agreement with previous theoretical analysis,[9] making it particularly appealing for nanoscale motion sensing and on-chip optical actuation in applications where strong yet broadband optomechanical coupling is required. Decreasing the gap increases optical propagation losses, as a larger fraction of the optical power travels inside the metal. If the beam length (optical travel distance) is also decreased, for each length there is a corresponding gap (Fig. 4a inset) such that the insertion loss (loss through an unactuated device) remains constant, e.g. at 1/e power (4.3 dB) with length scaling $\sim g^{0.8}$. The striking result shown in Fig. 4a,b is that if we scale down the GPPM dimensions in this way, we will maintain the phase modulation range without incurring a loss penalty while simultaneously reducing both the length and the gap, by an order of magnitude or more, as the calculated phase and excess loss vs. gap plots illustrate. In fact, the phase modulation range stays constant with miniaturization for a given optical loss. For gaps much smaller than the SP evanescent decay distance, universal scaling emerges between the phase shift and the excess loss such that they are linearly related regardless of the unmodulated gap dimension, e.g. as the gap is decreased to 72 % of the initial gap, the phase modulation stays constant at $\pi$ radians and there is a small excess loss of 0.8 dB (Fig. 4b inset) independent of the device scale.

For static MIM devices it has already been shown that the lateral dimensions can be scaled down together with the gap height and that low-loss coupling from larger mode waveguides suitable for long-distance signal transmission can be achieved.[4,5] Coupling efficiency > 70 % was demonstrated, coupling from a waveguide 500 nm wide x 200 nm high to an 80 nm x 17 nm waveguide using a 29° linear coupling taper.[5] As long as the device width is larger than the gap, the optical mode remains well-confined in the gap under the beam.[5] Our GPPM model is quantitatively valid for device widths larger than approximately half the GP wavelength (estimated as 710 nm in the experiment, $n_{eff} \approx 1.1$). As the effective index increases and GP wavelength decreases with the gap ($\approx$ 370 nm and $n_{eff} \approx 2.1$ for a 17 nm gap), the width of the device can be reduced further as well. By way of example, approximately 10x linear downscaling keeps losses near 5 dB in a broadband, *non-resonant* modulator with a footprint < 1 µm$^2$ (17 nm gap, 400 nm beam width and 2 µm beam length). In such a scaled GPPM the optomechanical modulation strength is increased approximately inversely with the gap to about 560 mrad/nm, despite the length decrease.

Importantly, the electrostatic actuation amplitude also scales favorably with miniaturization. Within the applicability of linear beam-bending theory without in-plane stress the shape of the deformation remains self-similar, and the same percentage gap actuation can be achieved with voltage that scales as $V^2 \sim g^3/L^4 t^3$ where $g$ is gap height, $L$ is beam length and $t$ is beam thickness (see Methods). Given the beam-length/gap-height combinations chosen according to the chosen scaling constraint of Fig. 4a inset, $V \sim t^{3/2}$ is constant at fixed beam thickness, and approximately independent of the beam length and gap height. If necessary, the beam thickness can easily be scaled down by a factor of 4 or more (while staying well above the optical skin depth of approximately 25 nm), reducing the actuation voltage below 1 V, to a level compatible with low voltage CMOS circuitry.

The inset of Fig. 1e shows that the realized GPPM has resonance frequency of 812 ± 6 KHz, an air-damping-dominated quality factor of 2.74 ± 0.14, and was actuated at a drive frequency up to 1 MHz. While we emphasize that very high modulation frequency is not required for the envisioned on chip optical switching and reconfiguration applications, the mechanical resonance frequency scales as $\sim t/L^2$ and is able to increase up to $\approx$ 100 MHz with fixed beam thickness and drive voltage. Furthermore, non-plasmonic nanomechanical cantilever devices of similar dimensions have been made operating up to 1 GHz with careful material choice,[30,31] and such a fast, yet ultrasmall modulator can potentially operate at low voltage with the use of piezoelectric actuation.[32-34]

Considering the negligible power dissipation of its electrostatic-drive, actuation voltages at the level of the smallest high speed transistors, its length scale and feature size at the level of CMOS metallization layers, its broadband optical operation and its reasonable speed, we argue that a GPPM can play a unique and important role as a building block for optoelectronic integration. A device with these features is particularly well suited as an element for on chip reconfigurable switch fabrics for future dynamic inter- and intra-chip optical communication architectures.

Beyond conventional 1x2 and 2x2 switches, it is possible to array such devices side-by-side to form spatial plasmon modulators and implement, for example, single-stage 1xN switching and arbitrary multiport beam splitting: functionalities demonstrated using spatial light modulators in free space. Reconfigurable routing of photonic signals or reconfigurable flat plasmonic optics, where *local* phase modulation across an extended GP wavefront could be used to shape, focus or guide GP propagation via *independent* actuation of multiple adjacent modulators. The authors chose a multiple-beam GPPM as a step in that direction when a single beam device would have sufficed.

In summary we have experimentally demonstrated exceptionally strong optomechanical transduction with low optical losses in electrostatically actuated nanoscale gap MIM plasmon modulators. The 23 μm long GPPM with 52 mrad/nm optomechanical modulation strength at 780 nm achieved 5 radians of phase modulation with low insertion and excess losses. An analytical model in good agreement with the measurements argues for direct miniaturization of these devices to sub 1 μm$^2$ footprint without degradation in optical performance and with an increase in speed and decrease in actuation voltage. This new concept enables a new class of on chip optical switching and optical circuit reconfiguration functionality.

**Methods**

**Nano-fabrication and operation.** A gold-SiO$_2$-gold stack with sputtered Au and PECVD (plasma enhanced chemical vapor deposition) SiO$_2$ layers, all three 220 ± 5 nm thick, was deposited onto nominal 500 µm thick borosilicate glass with ≈ 10 nm Cr adhesion layer located between the substrate and bottom Au layer and ≈ 2 nm thick Ti adhesion layer on both sides of the SiO$_2$. All device features, except the out-coupler slit, were lithographically written with EBL (e-beam lithography) using ≈ 500 nm PMMA (poly methyl methacrylate) e-beam resist. After resist development, device components were Ar ion milled into the top Au layer. The beams were released by wet etching of the underlying SiO$_2$ in 6:1 BOE (buffered oxide etch) with subsequent CO$_2$ critical-point drying. The SiO$_2$ was completely removed everywhere below the lithographic patterns leaving a lateral undercut of ≈ 2.5 µm. After release, the SiO$_2$ pillars supporting the beams at their ends were ≈ 3 µm wide in the direction of GP propagation. The out-coupler slits were ≈ 150 nm wide by ≈ 20 µm long and were cut with a focused ion beam (FIB). The suspended in-coupler gratings composed of strips ≈ 18 µm long and ≈ 400 nm wide with periods of ≈ 720 nm and ≈ 760 nm were electrically grounded to avoid unintended actuation.

An electrically isolating 2 µm wide trench in the top Au layer surrounded the GPPM components (partial-view in Fig. 1b). A narrow wire nano-fuse connected the area inside the trench to that outside to allow charging to dissipate during scanning electron microscopy and FIB. The nano-fuses were electrically severed before electrostatic beam actuation. The actuation voltage between the beams and the bottom gold film was applied via probes electrically connected to the top and bottom films.

The GP-only intensity profiles seen in Fig. 2a exhibit some non-Gaussian that we attribute to small fluctuations in the heights (gaps) of individual beams when stress was relieved in the top gold and PMMA during release. This effect varied from device to device and can be seen in Fig. 3b the intensity variations as the gap narrowed.

**Interferometer.** A Mach-Zehnder-type interferometer was used to measure the phase shift between a GP and a reference laser beam. It consists of an inverted microscope custom fitted with a top excitation objective and beam steering optics. Laser light ($\lambda$ = 780 nm, linewidth < 200 kHz and power ≈ 0.2 mW) was fiber-coupled to the top, collimated to about 1.5 mm, and incident upon a 50/50 beamsplitting cube. Half of the light was directed to the objective while the other half formed a reference beam, circling back on itself using adjustable mirrors over a ≈ 20 cm path length before also travelling into the 10x excitation objective. The excitation beam, an 8 μm diameter focused spot, was placed onto the in-coupler grating, directly launching a GP through the waveguide. The reference beam was focused onto the out-coupler slit at a 13.2 ± 0.05° angle with respect to the normal. Near the out-coupler slit the top gold film was removed and the GP from the device continued to propagate as an SP on the gold-air interface. At the slit, the reference beam interference with the propagating SP developed into fringes. Gap narrowing by electrostatic beam actuation causes GP phase velocity retardation, and thus shifts the interference fringes compared with their initial positions. The angled reference beam was chosen to show multiple interference fringes across the out-coupler slit. For each device the reference beam intensity is adjusted to maximize interference visibility before voltages are applied.

The reference incidence angle was measured with no device in place by analyzing a series of images of the reference laser spot as it moved across the microscope objective focal plane as the objective was translated vertically by a known amount.

**Dynamics**. The mechanical response of the GPPM was measured with a strobed white-light optical profiler using harmonic small positive actuation voltages up to 1 MHz. The strobed pulses were phase delayed for a phase sensitive motion measurement. The response amplitude can be seen in Fig. 1e where it is fit by a dampened harmonic oscillator model.

**Gaussian fits.** The phase difference between the GP and reference laser was extracted from Gaussian interference fits of the measured interference profiles seen in Fig. 2b and is the only variable used. The fits use: 1) GP-only intensity profile data like that seen in Fig. 2a; 2) Gaussian reference beam parameters (width, peak position and integrated area) extracted from measured interference profiles with the reference beam intensity maximized; and 3) independently measured reference incidence angle (Fig. 2a). Fits from the data of one device can be seen plotted in Fig. 2b. The results in Fig. 3 from several devices are the same within the experimental error when adjusted for the initial phase differences caused by slightly different unactuated gaps.

**Uncertainty.** The uncertainties reported throughout the manuscript represent one standard deviation statistical uncertainties, unless otherwise indicated. The uncertainty in device sizes is given by a conservative estimate of the scale calibration accuracy of the electron microscope used. The uncertainties in the gap values are the standard deviations of gaps under different beams in a single device under a given applied voltage, likely due to mechanical variations from beam to beam. The measurement imprecision and errors in actuation repeatability are much smaller. The uncertainty in the phase measurement is dominated by the slow drift of the optical wavelength, which results in a phase drift between excitation and reference beams going through unequal paths. Therefore we make separate reference unactuated phase measurements before and after each nonzero voltage phase measurement. We use the variation in the unactuated measurements to establish the statistical phase measurement uncertainty reported, while the statistical uncertainty of the fitting procedure for each individual interferogram is much smaller.

**Theory.** To theoretically understand the GPPM performance, we developed a semi-analytical model of one dimensional GP propagation, assuming an infinitely-wide plane wave GP, quadratic beam profiles, semi-infinite MIM gold layers and vacuum in the gap. The device was broken into 1 nm intervals in the direction of GP propagation with each interval assigned a gap-dependent effective refractive index and the corresponding wavenumber. Using continuity boundary conditions from Maxwell's equations, the phase shift and intensity was cumulatively calculated. The analytic results of GP phase shift and intensity calculations agree well with measured data (Fig. 3a, b solid line) [Supplementary]. While the modeling procedure includes both forward and backward propagating waves, under the experimental conditions the gap changes adiabatically and the back-propagating power was found to be negligible throughout the model

**Mechanics.** The electrostatic pressure $P$ at the cantilever bottom is proportional to $P \sim (V/g)^2$. Within the applicability of linear elastic beam bending theory without in-plane stress the shape of the deformation remains self-similar with size scaling, and the magnitude is proportional to $z \sim P \cdot L^4/t^3$, thus if we require $z \sim g$, then $g \sim (V/g)^2 \cdot L^4/t^3$ or $V^2 \sim g^3 t^3/L^4$.

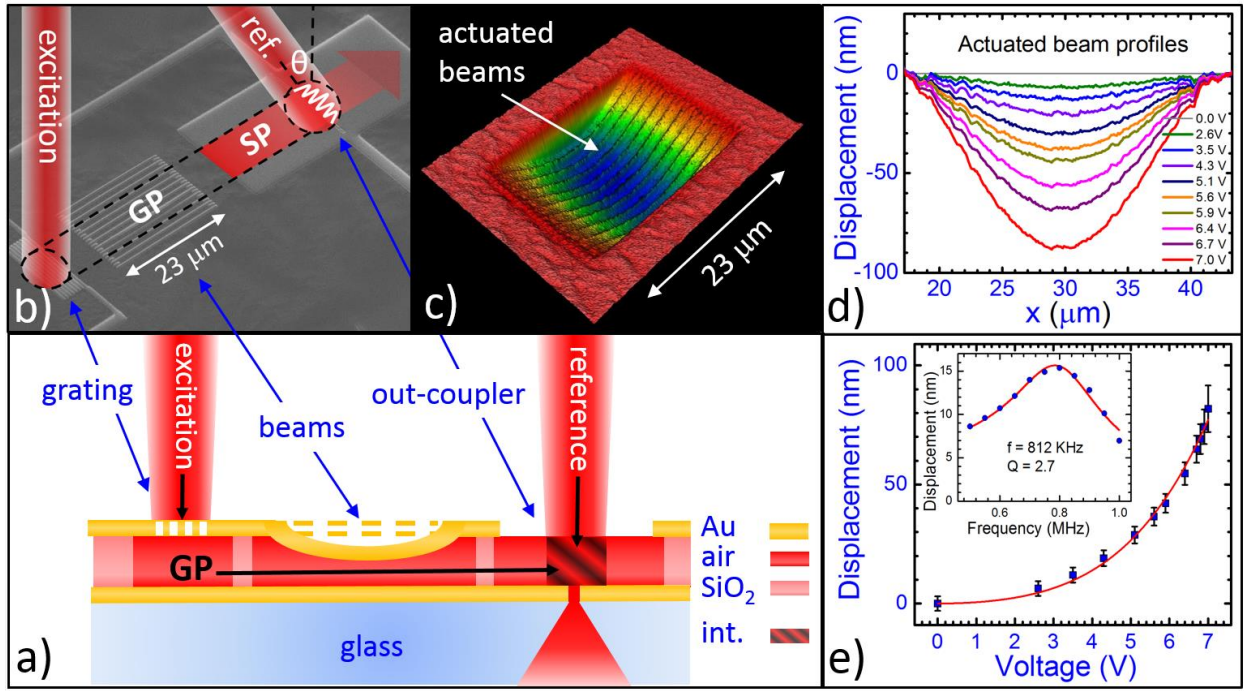

**Figure 1 | GPPM. a,** Schematic: MIM gap plasmons are directly launched via grating coupling with a focused free-space excitation laser, propagate under the gold beams and exit as surface plasmons[27] to the bottom gold/air interface. A focused reference beam, split from the excitation laser and incident at 13.2°, interferes with the plasmon at the out-coupler slit. Light is collected from below and imaged onto a camera. Electrostatic actuation of the beams towards the substrate phase-retards the GP. **b,** Scanning electron micrograph of the GPPM with cartoon overlays of the excitation and tilted reference lasers, propagating GP and interference fringes at the out-coupler. The 11 beams are 1.5 μm wide separated by 150 nm. **c,** Interferometric micrograph showing GPPM beams actuated towards the substrate with 6.5 V. Depth exaggerated and color coded for clarity. **d,** Electrostatically actuated beam displacement profiles. **e,** Beam displacement at the middle vs. actuation voltage. The solid red line is a guide for the eye. **Inset:** Beam's displacement amplitude vs. the frequency of applied harmonic electrostatic excitation. Line is a simple harmonic oscillator fit.

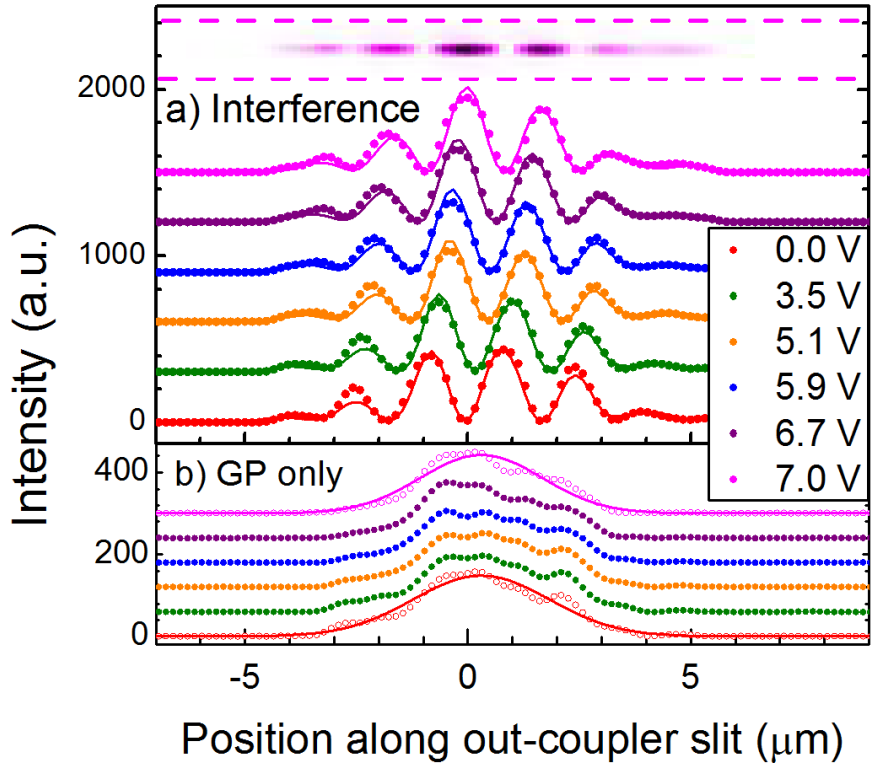

**Figure 2 | Measured out-coupler intensity profiles. a,** Evolution of the interference fringes with actuation voltage. Solid lines are interference fits with varying phase. Note that peaks at 0.0 V (red) become troughs at 7.0 V (magenta). **Inset (top):** Color mapped image of the interference fringes seen at the out-coupler slit at 7.0 V. The dark magenta regions correspond to the interference peaks directly below. The intensity profiles were obtained by vertical integration of the pixel intensities between the dashed lines. **b,** GP intensity without the reference beam vs. relative position across the out-coupler slit. Solid lines are Gaussian fits. All plots in this figure are vertically shifted for clarity.

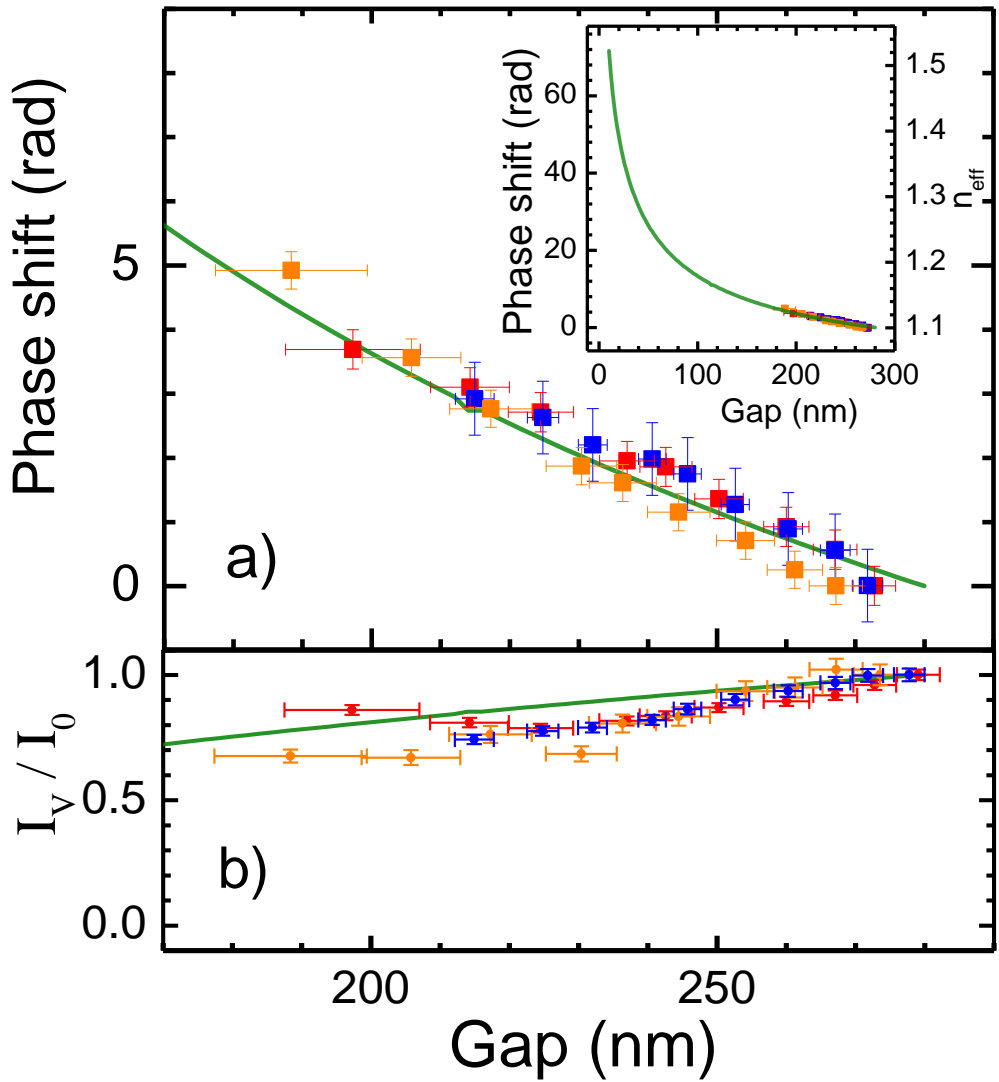

**Figure 3. GPPM phase modulation and excess optical loss | a,** Plasmon phase shift ΔΦ relative to the unactuated device state vs. gap. Green line is a calculation and points are measured (four devices are the same within the experimental error when adjusted for the initial phase differences caused by slightly different unactuated gaps). Gap is the minimum gap created when a beam is actuated. The vertical error bars are ± the standard deviation due to random interferometer phase drift. Horizontal error bars are ± the standard deviation of the displacement of

multiple individual beams. **Inset,** Extended range plot showing increased phase modulation strength and effective index tuning at small gaps. **b,** Integrated GP intensity $I_v$ vs. gap from Gaussian fits of several devices. Analytical (green line) and data points are normalized to the unactuated GP intensity $I_0$. Vertical error bars are ± the standard error from Gaussian fits of intensity profiles.

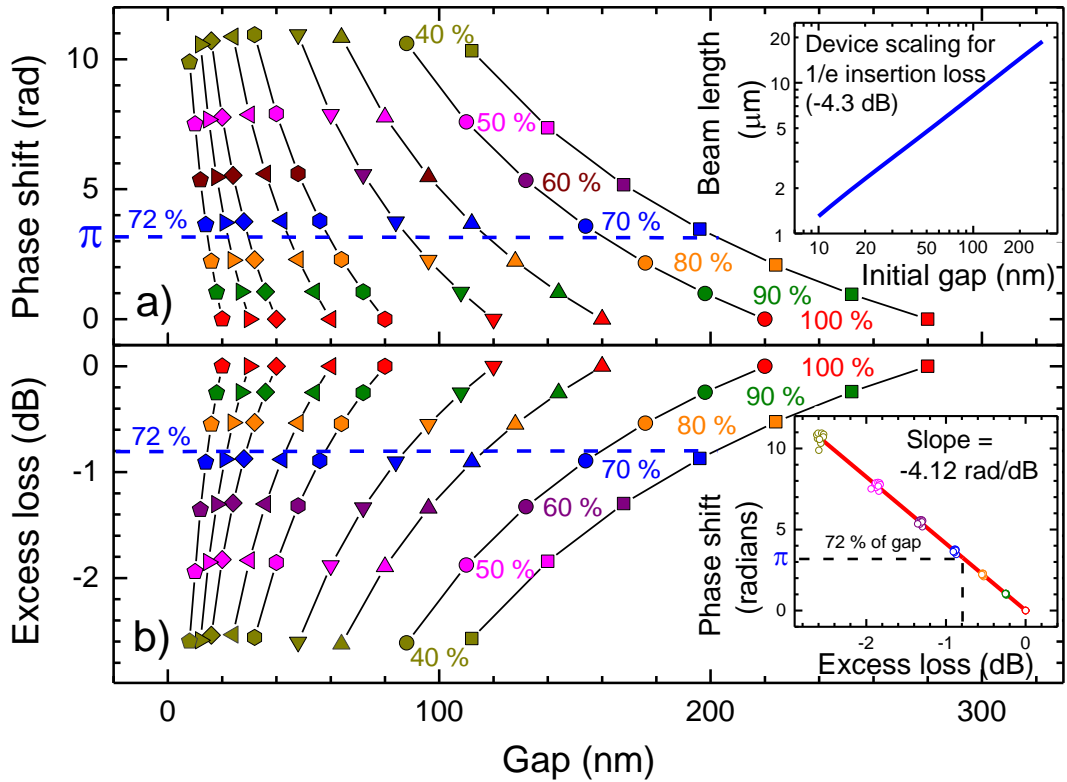

**Figure 4. GPPM Scaling | a,** Calculated phase shifts vs. gap for initial gaps varying from 280 nm down to 20 nm. The beam length of each line is chosen to give a 1/e (-4.3 dB) insertion loss for a given initial gap (inset). The lines show how the phase changes as the gap is actuated down from 100 % to 40 % of the initial gaps. Identically-colored points indicate the same percent beam actuation. An actuation depth of ≈ 72 % results in a phase shift of π radians (dashed line). Regardless of the initial gap, for the same percent actuation depth, the phase shift is almost constant as indicated by the horizontal rows of identically-colored points. **b,** Calculated excess loss vs. gap using the same beam lengths described. Excess loss is defined relative to the unactuated state. The same universality is seen here; e.g. ≈ 72 % actuation of the initial gap gives ≈ 0.8 dB loss (dashed line). **Inset:** Phase shift vs. excess loss is linear and independent of device scale.

**Acknowledgements**

The authors acknowledge support from the Measurement Science and Engineering Research Grant Program of the National Institute of Standards and Technology and AFOSR under Grant No. FA9550-09-1-0698. The authors thank Dr. Amit Agrawal and Dr. Henri Lezec for their technical suggestions and insightful comments on the manuscript, Glenn Holland and Alan Band for their technical help with the experimental setup and P. Lubik for his programming assistance. Computational support of the Department of Defense High Perfomance Computation Modernization project is acknowledged. This work was performed, in part, at the Center for Nanoscale Materials, a U.S. Department of Energy, Office of Science, Office of Basic Energy Sciences User Facility under Contract No. DE-AC02-06CH11357.

**Author contributions**

B.S.D. developed the fabrication process, designed and fabricated the modulators, performed the experiments, analyzed the data and wrote the manuscript. M.I.H. developed an analytical model and wrote the manuscript. G.B. developed the concept, designed the experiment and wrote the manuscript. D.A.C and D.L. developed the fabrication process, V.A. developed the concept, designed the experiment, performed simulations, developed the fabrication process, analyzed the data and wrote the manuscript.

## Additional Information

Supplementary information is available in the online version of the paper. Reprints and permissions information is available online at www.nature.com/reprints. Correspondence and requests for materials should be addressed to V.A.

## Competing financial interests

The authors declare no competing financial interests.

# Supplementary Information

# Ultracompact Nano-Mechanical Plasmonic Phase Modulators

B. S. Dennis[1], M. Haftel[2], G. Blumberg[1], D. A. Czaplewski[3], D. Lopez[3] and V. Aksyuk[4#]

[1]Rutgers, the State University of New Jersey, Dept. of Astronomy and Physics, Piscataway, NJ 08854, USA, [2]University of Colorado, Dept. of Physics, Colorado Springs, CO 80918, USA, [3]Argonne National Laboratory, Center for Nanoscale Materials, Argonne, IL 60439, [4]Center for Nanoscale Science and Technology, National Institute of Science and Technology, Gaithersburg, MD 20899, USA. [#]e-mail: vladimir.aksyuk@nist.gov

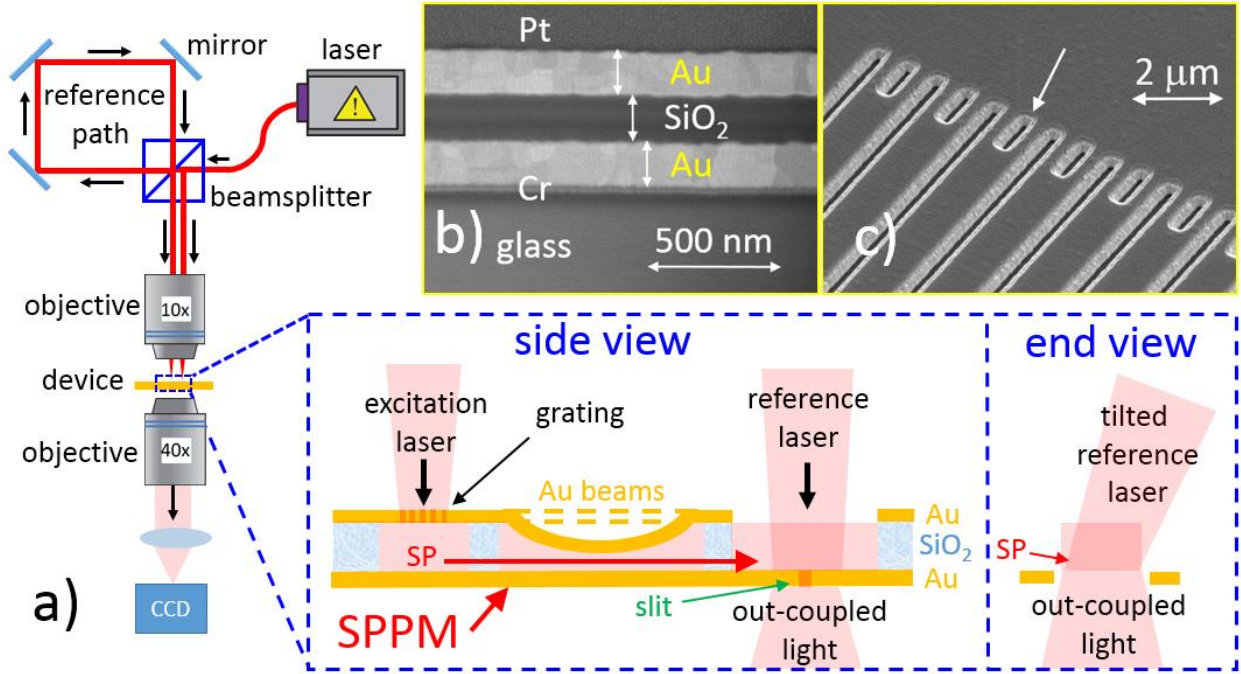

**Supplementary Figure 1 | Mach Zehnder-type interferometer and GPPM. a,** Laser light was coupled to an inverted microscope fitted with a custom top objective. A beamsplitter directed the excitation laser to an in-coupler grating and the reference laser to the out-coupler slit. Side view of the GPPM shows an MIM plasmon launched into and propagate through the actuated waveguide and interfere with the reference laser at the out-coupler slit. Beams can be pulled down to one third of the gap before instability occurs. End view shows the tilted reference laser in a cross section at the out-coupler slit. **b,** electron micrograph of the Au/SiO$_2$/Au MIM cross-section. Vertical arrows are 220 nm. Pt layer added for protection during cross-sectioning. **c,** All beam ends had an additional 3 nm long medial slit (white arrow) that aided in smoothing the optical surfaces of the SiO$_2$ pillars during release.

## 1. GPPM Analytical Model

To estimate the transmission and phase of a GP propagating in a GPPM (Fig. 1), a model is chosen that consisting of an MIM slot waveguide with a bottom flat gold surface, a gold beam suspended above the bottom gold surface, and vacuum in the gap between the bottom surface and the beam. The thicknesses of the bottom layer and beam are taken to be semi-infinite, which is a good approximation if the thicknesses are much greater than the skin depth ($\approx$ 25 nm). In the results that follow, we use constant heights and parabolic profiles, fit to the experimental devices. The vacuum gap has a height profile $h(x)$ described by

$$h(x) = h_0 \quad (x < 0 \text{ or } x > L) \tag{1a}$$

$$h(x) = ax^2 + bx + c \quad (0 < x < L), \tag{1b}$$

where $a$, $b$, and $c$ are such that the parabolic form give $h = h_0$ at $x = 0$ and $x = L$ with a minimum height $h_{min}$ at x = $L/2$, and $L$ is the length of the parabolic region. Given that the GP propagating in the waveguide is incident at $x = 0$, the transmission and phase as it exits the nonconstant region at $x = L$ (and the reflection at $x = 0$) are determined.

To estimate the GP field, GPs propagating in waveguides like those above with a constant height profile $h(x) = h_0$ for various gaps $h_0$ are first considered. For such an infinitely long waveguide the spatial GP fields are given by

$$H_y(x,z) = exp(ik_x x) [A_i \, exp(ik_{zi} z) + B_i \, exp(-ik_{zi} z)] \tag{2a}$$

$$E_x(x,z) = (k_{zi}/k_0) \; exp(ik_x x) \; [A_i \; exp(ik_{zi}z) - B_i \; exp(-ik_{zi}z)] \quad (2b)$$

$$E_z(x,z) = (-k_x/k_0) \; exp(ik_x x) \; [A_i \; exp(ik_{zi}z) + B_i \; exp(-ik_{zi}z)], \quad (2c)$$

where *i* labels the layer (1 for bottom Au, 2 for vacuum, 3 for top Au) and

$$k_x^2 + k_{zi}^2 = k_0^2 \varepsilon_i. \quad (2d)$$

In Eq. (2), $k_0 = \omega/c$, $x$ is the coordinate along the waveguide, $z$ is the direction normal to the surface and $\varepsilon_i$ is the dielectric constant of layer $i$. The coefficients $A$ and $B$ are determined in each layer by the continuity boundary conditions of the tangential components $H_y$ and $E_x$ at the two gold-vacuum interfaces separated by the distance $h_0$. For a GP mode the field must be bounded in the $z$ direction, i.e., the decay is in the negative $z$ direction in layer one and in the positive $z$ direction in layer three. This means $A_1 = B_3 = 0$ and $k_{z1}$ and $k_{z3}$ have positive imaginary parts. This condition is satisfied only for discrete values of $k_x$, corresponding to guided modes. All modes but one have a low frequency cutoff, and for small vacuum gaps in gold, in the visible and near infrared, only one complex value of $k_x$ satisfies this condition and is propagating ($Re(k_x) >> Im(k_x)$). The wave number $k_x$, which depends on $h_0$, is denoted as $k_{GP}(h_0)$.

The strategy is to utilize a spatially dependent wave number $k_x(x)$, equated to $k_{GP}(h(x))$, in integrating the fields in the $x$ variable. Since only a constant $k_x(x)$ can be integrated analytically, we discretize the parabolic region of the $x$ axis into $N$ equally spaced intervals and set the wave number $k_x$ in each interval to a constant value $k_{xi}$ appropriate to the height at the midpoint of the interval, $i$, i.e.,

$$k_{xi} = k_{GP}(h(x_i)), \quad (3)$$

where $x_i$ is the midpoint of interval $i$. From this interval-dependent wave number, we define an effective dielectric constant $\varepsilon_{\text{eff }i}$ in interval $i$

$$\varepsilon_{\text{eff }i} = k_{xi}^2 / k_0^2. \tag{4}$$

The accuracy of a discretization method increases as the intervals become smaller. Also, the discretization is more accurate if $k_x(x)$ is more slowly varying with $x$. The actual GPPM beam height profiles only dip several tens of nm over beam lengths measured in μm. Intervals of 50 nm gave results that differed from using 10 nm intervals by about 1 %. Actual calculations used 1 nm intervals.

The final approximation replaces the GP with a plane wave normally incident on a stack of dielectric intervals with dielectric constants given by Eq. (4) (or wave numbers given by Eq. (3)). Specifically the fields to be integrated have the form in interval $i$

$$H_y^{\text{eff}}(x) = C_i \exp(ik_{xi}x) + D_i \exp(-ik_{xi}x) \tag{5a}$$

$$E_z^{\text{eff}}(x) = (-k_{xi}/k_0) [C_i \exp(ik_{xi}x) - D_i \exp(-ik_{xi}x)] \tag{5b}$$

where $C$ and $D$ are coefficients determined by demanding continuity at each interval boundary, subject to the outgoing wave boundary condition outside the last boundary and an incident plane wave before the first, i.e.,

$$C_0 = 1 \tag{6a}$$

$$D_{N+1} = 0. \tag{6b}$$

where N is the total number of intervals. The reflection coefficient is then simply $r = |D_0|^2$ and the transmission coefficient $t = |C_{N+1}|^2$. One way to integrate Eq. (5) is to start with $C_{N+1} = 1$, $D_{N+1} = 0$, then determine $C_i$ and $D_i$ in each interval, proceeding backwards starting from $i = N$ to $i = 0$, by demanding continuity in the fields of Eqs. (5a) and (5b) at the boundary of intervals $i$ and $i + 1$. Doing this yields a value of $C_0$ generally not equal to unity. Since all equations are linear, all the $C_i$ and $D_i$ are renormalized such that Eq. (6a) is satisfied. This method avoids matrix operations. In addition to transmission and reflection coefficients, the full fields inside the parabolic region and on exiting are calculated, which means the amplitude and phase can be tracked throughout. The main interest is in the phase shift $\Delta\phi$ and transmission intensity $I$ as the GP exits the parabolic (or nonconstant) region of length $L$, where the phase shift is the difference in phase between the exiting wave and that which would occur for a wave exiting this region when the height profile is a constant $h(x) = h_0$. These quantities are given by

$$I = |C_{N+1}|^2 \exp(-2\ Im(k_{GP}(h_0)L) \tag{7}$$

$$\Delta\phi = Arg(C_{N+1}) \tag{8}$$

The exponential factor in Eq. (7) occurs because the GP wave number is generally complex and because the dielectric constant of gold is complex. Thus, even for a waveguide with planar surfaces separated by a constant height $h_0$, losses determined by the exponential factor occur. Eq. (8) determines the phase shift only to an integral multiple of $2\pi$ since the *Arg* function only takes on values between $-\pi$ and $\pi$ (In the final analysis, the $2n\pi$ differences have no physical consequences for interference effects). To get the proper phase shift one needs to examine the phase shift at the end of each interval and whenever a jump of (or close to) $\pm 2\pi$ occurs add (+ or -) $2\pi$ to the cumulative phase shift to make it (nearly) continuous. The overall phase shift is

$$\Delta\phi = \Sigma_{i=1,n}\, \Delta\phi_i$$

where $\Delta\phi_i$ is the phase shift in the interval i.

$$\Delta\phi_i = Arg(\, H_y(x_i)/[\, H_y(x_{i-1})\, exp(k_{GP}(h_0)(x_i - x_{i-1})]) + 2m_i\pi, \qquad (9)$$

where $x_i = i\, L/N$ and $m_i = -1, 0,$ or $1$ if $\Delta\phi_i \sim 2\pi, 0, -2\pi$, respectively, i.e., $m_i$ is such as to keep the cumulative phase shift continuous whenever there is a $\pm 2\pi$ jump from the Arg function.

In Eq. (5) the $z$-dependence of the SP fields is suppressed, as well as the longitudinal $E_x$ component, which depends on the $z$-dependence of the magnetic field. One physical argument for this is that the energy transfer in the SP is determined by the real Poynting vector $S_y$, and this does not depend on the longitudinal component of the electric field. As for the $z$-dependence, it is not explicitly involved in Eq. (5), but implicitly comes into play in Eq.(2) as integration of this equation in the $z$ direction determines $k_{GP}(h_0)$ which eventually determines $k_{xi}$ in Eq.(5). The local approximation for the longitudinal wave number (Eq. (3)) depends (much like an adiabatic approximation) on a slow variation in $h(x)$, which largely holds for the devices considered.

The gold dielectric constant we use in calculating $k_{GP}(h_0)$ from Eq. (2) is determined by an analytic fit to the frequency dependence of the values measured by Johnson and Christy.[1] In particular, for the wavelength of interest of 780 nm, the dielectric constant of gold used is *ε(Au-780 nm) = 22.4476 + 1.36505 i*. The beam is divided into 1.0 nm layered segments and the boundary condition matching of $E^{eff}$ and $H^{eff}$ is performed at each interface as in the discussion of Eq. (5). The resulting GP phase shifts and intensities throughout are in good agreement with experiment.

**References.**